\documentclass[12pt]{article}

\usepackage[utf8]{inputenc}
\usepackage{amsmath,amsfonts,graphicx}
\usepackage[square,numbers,sort&compress]{natbib}
\usepackage{fullpage}
\usepackage{lineno}
\usepackage{xcolor}
\usepackage{pdflscape}
\usepackage{hyperref}
\usepackage{authblk}
\usepackage{multicol}

\newcommand{\highlight}{}

\title{Estimation of end-of-outbreak probabilities in the presence of delayed and incomplete case reporting}

\author[1]{M. J. Plank*}
\author[2]{W. S. Hart\textsuperscript{\textdagger}}
\author[3]{J. Polonsky}
\author[4,5]{M. Keita}
\author[6]{S. Ahuka-Mundeke}
\author[2]{R. N. Thompson\textsuperscript{\textdagger}}

\affil[1]{School of Mathematics and Statistics, University of Canterbury, Christchurch, New Zealand}
\affil[2]{Mathematical Institute, University of Oxford, Oxford, UK}
\affil[3]{Geneva Centre of Humanitarian Studies, University of Geneva, Geneva, Switzerland}
\affil[4]{World Health Organization, Regional Office for Africa, Brazzaville, Democratic Republic of the Congo}
\affil[5]{Institute of Global Health, Faculty of Medicine, University of Geneva, Geneva, Switzerland}
\affil[6]{National Institute of Biomedical Research, Kinshasa, Democratic Republic of the Congo}

\footnotetext{Author for correspondence: michael.plank@canterbury.ac.nz}
\footnotetext{These authors contributed equally to this work.}

\date{}

\begin{document}


\maketitle


\begin{abstract}

Towards the end of an infectious disease outbreak, when a period has elapsed without new case notifications, a key question for public health policy makers is whether the outbreak can be declared over. This requires the benefits of a declaration (e.g., relaxation of outbreak control measures) to be balanced against the risk of a resurgence in cases. To support this decision making, mathematical methods have been developed to quantify the end-of-outbreak probability. Here, we propose a new approach to this problem that accounts for a range of features of real-world outbreaks, specifically: (i) incomplete case ascertainment; (ii) reporting delays; (iii) individual heterogeneity in transmissibility; and (iv) whether cases were imported or infected locally. We showcase our approach using two case studies: Covid-19 in New Zealand in 2020, and Ebola virus disease in the Democratic Republic of the Congo in 2018. In these examples, we found that the date when the estimated probability of no future infections reached 95\% was relatively consistent across a range of modelling assumptions. This suggests that our modelling framework can generate robust quantitative estimates that can be used by policy advisors, alongside other sources of evidence, to inform end-of-outbreak declarations.

\end{abstract}


\newpage

\section*{Introduction}

Infectious disease outbreaks are responsible for devastating consequences. This is not only because of the negative consequences of infection (such as, for severe cases, hospitalisation or death), but also because public health measures introduced to counter outbreaks are costly and can place a substantial burden on the host population \cite{tildesley2022optimal,dobson2023balancing}. Determining when an outbreak can be declared over and interventions can be relaxed or removed is therefore an important question with wide-reaching health and economic ramifications \cite{klepac2015six}. 

A ``rule of thumb'' used by the World Health Organization for diseases such as Ebola virus disease (EVD) \cite{coltart2017ebola} and Marburg virus disease \cite{who2023marburg} is that outbreaks can be declared over following two maximal incubation periods without cases. A similar approach has been used for Covid-19 in countries such as New Zealand that temporarily eliminated community transmission in 2020 and 2021 \cite{baker2020successful}. The rationale for this guideline is that it might be expected that missing two successive generations of cases is unlikely. However, such a rule ignores the characteristics of the specific outbreak under consideration, with factors such as the reproduction number and case ascertainment rate \cite{thompson2019rigorous,parag2020exact,djaafara2021quantitative,thompson2024using} expected to affect the risk that an outbreak is over after a fixed period with no newly notified cases.

In recent methodological work, epidemiological modellers have developed approaches for determining when specific infectious disease outbreaks can be declared over with a certain level of confidence \cite{nishiura2016objective,nishiura2016methods,hart2024optimizing,bradbury2023exact,linton2022epidemics}. These methods have centred around estimation of the {\em end-of-outbreak probability}, meaning the probability that no further cases will occur after the current date. When the end-of-outbreak probability reaches a high value, then in principle a more evidence-informed and outbreak-specific end-of-outbreak declaration can be made compared to declaring an outbreak over after a pre-determined period without cases.

While recent methods for estimating the end-of-outbreak probability have involved substantial theoretical advances \cite{bradbury2023exact,hart2024optimizing}, a key omission from those approaches is the fact that reporting delays and underascertainment of cases have not been accounted for in a rigorous and flexible fashion. For example, Djafaraa et al. \cite{djaafara2021quantitative} and Thompson et al. \cite{thompson2024using} included case underascertainment in some analyses, but both did so in a simple fashion by estimating the relative likelihood of unreported cases arising at different times based only on the timing of reported cases. In reality, the number of unreported cases on a given day depends on the timing of all other cases (reported cases and unreported cases), since all previous cases act as potential infectors and all subsequent cases act as potential infectees. Since both reporting delays and case underascertainment beset inference of key epidemiological quantities, they would be expected to affect estimates of the end-of-outbreak probability. Hence, rigorous inclusion of these factors in models for inferring the end-of-outbreak probability and associated quantities is extremely important.

 Heterogeneous transmission patterns are known to increase the extinction probability for an outbreak starting with a single infected individual because, for a given reproduction number, there is a higher likelihood that a case will not generate any secondary infections \cite{lloyd2005superspreading}. For a similar reason, two recent studies found that a higher degree of heterogeneity can lead to higher end-of-outbreak probability values \cite{bradbury2023exact,hart2024optimizing}. However, Thompson et al. \cite{thompson2024using} did not obtain higher end-of-outbreak estimates when accounting for heterogeneity in the total number of daily new infections (which is slightly different to heterogeneity in the number of secondary infections caused by individuals over their whole infectious period), as a consequence of such heterogeneity affecting reproduction number estimates used in end-of-outbreak probability calculations. Similarly, distinguishing between imported and locally acquired cases is known to affect estimates of the effective reproduction number \cite{thompson2019improved}, which will in turn affect estimates of the end-of-outbreak probability.

In this manuscript, we combine the method of Thompson et al. \cite{thompson2024using} for estimating end-of-outbreak probabilities from a time series of case incidence data, with a hidden-state model that provides greater flexibility to model relevant processes and variables \cite{abbott2020estimating}. The hidden-state model is based on the renewal equation, which provides a popular framework for estimation of the instantaneous reproduction number and near-term epidemic forecasting \cite{cori2013new,thompson2019improved,watson2024jointly}. The model treats the instantaneous reproduction number and the daily incidence of new infections as hidden states and daily case notifications as an observed state. We estimate the hidden states from observed data using a particle filter method.

We show how this framework can be used to estimate end-of-outbreak probabilities in the presence of underascertainment of cases, distributed lags from infection to case notification, heterogeneous transmission patterns, and data that classify cases as either imported or locally acquired. We also consider alternative ways in which the end of an outbreak may be determined, by investigating three different possible definitions of the end of an outbreak, specifically: (i) no future infections; (ii) no future infections or notifications; or (iii) no sustained future chains of transmission.

We apply our model to case studies from two very different outbreaks: an outbreak of EVD in \'Equateur Province, Democratic Republic of the Congo (DRC), in 2018; and an outbreak of Covid-19 in New Zealand in 2020. Both outbreaks were ultimately declared over, initiating the relaxation of public health interventions \cite{who2018ebola,baker2020successful}. We investigate how the estimated end-of-outbreak probabilities depend on model parameters, and how the dates on which a threshold estimated end-of-outbreak probability is reached compare to the actual dates on which the respective outbreaks were declared over.

\section*{Methods}

Here, we describe the transmission model underlying our analyses, the inference procedure that we implemented and our approaches for estimating the end-of-outbreak probability. We also describe the two outbreak datasets that we analysed (Covid-19 in New Zealand and EVD in DRC). Data and code to run the model and reproduce the results in this article are publicly available at \url{https://github.com/michaelplanknz/end_of_outbreak_analysis}.

\subsection*{Transmission model}

We use a hidden state model in which the instantaneous reproduction number $R_t$ at time $t$, the daily incidence of new local infections $I_t$ and related quantities are treated as hidden states. The number of new case notifications $C_t$ each day is treated as an observed state.

The instantaneous reproduction number $R_t$ is assumed to follow a Gaussian random walk in logarithmic space:
\begin{equation} \label{eq:Rt}
   \ln( R_t) \sim N\left( \ln(R_{t-1}) + \delta_{t-1} , \sigma_{t-1}\right),
\end{equation}
where $\delta_t$ and $\sigma_t$ are the mean and standard deviation of the random walk step on day $t$. We set $\delta_t=0$ and $\sigma_t$ to be a constant for most of the simulated time period. However, in the two case studies that we consider, in a short time window around the start of the interventions, we set $\delta_t<0$. This allows $R_t$ to decrease rapidly in response to the interventions.

To allow for individual heterogeneity in transmission rates, we assume that each infected individual is assigned a ``transmission multiplier'', representing variation in infectiousness and/or contact rates between individuals. We assume the transmission multipliers are independent, identically distributed random deviates from a Gamma distribution with shape $k_r$ and scale $1/k_r$ (i.e. mean $1$). Thus the total of the transmission multipliers for people infected on day $t$ is  
\begin{equation} \label{eq:Yt}
    Y_t \sim \Gamma\left( k_r\left( I_t+ \varepsilon n_t\right), 1/k_r   \right),
\end{equation}
where $n_t$ is the number of imported infections arising on day $t$ and $\varepsilon$ is the average infectivity of imported infections relative to local infections. The variable $Y_t$ represents the aggregate infectivity of people who were infected on day $t$.

The number of new local infections on day $t$ follows a standard renewal equation \cite{cori2013new} but with the aggregate infectivity of infections $s$ days previously ($Y_{t-s}$) used in place of the number of infections $s$ days previously ($I_{t-s}$) to drive the number of local infections on day $t$:
\begin{equation} \label{eq:renewal}
    I_t \sim \mathrm{Poiss}\left( R_t \sum_{s=1}^{t-1} Y_{t-s}  g_s  \right),
\end{equation}
where $g_s$ is the probability mass function for the generation interval distribution. Under this formulation, if the reproduction number was a constant $R$, then the total number of people infected by a randomly selected infected individual would follow a negative binomial distribution with mean $R$ and dispersion parameter $k_r$ \cite{lloyd2005superspreading}. In the limit $k_r\to \infty$, $Y_t$ is deterministically equal to $I_t+\varepsilon n_t$, meaning that Eq. \eqref{eq:renewal} reduces to the standard Poisson renewal equation. Smaller values of $k_r$ correspond to larger variance of $Y_t$, representing greater individual heterogeneity in transmission rates. 

{\highlight Note that the formulation of Eq. \eqref{eq:renewal} relies on the assumption that the number of secondary infections arising from each infected individual is independent of other infected individuals. This means that, although the model includes individual heterogeneity, it does not include network-type effects such as a higher probability of highly connected individuals infecting other highly connected individuals. Furthermore, Eq. \eqref{eq:renewal} assumes that the generation interval distribution is known. This is a standard modelling assumption, but we caution that estimating the generation interval distribution from epidemiological data is subject to various biases, particularly during the exponential growth phase of a newly emerging outbreak \cite{britton2019estimation,charniga2024best}.  }

To account for delays between infection and case notification, we assign each infected individual a notification time (regardless of whether or not they are actually notified as cases) according to an infection-to-notification distribution with probability mass function $u$. Among individuals infected on a given day $s$, the number of these individuals $Z_{st}$ with a notification time on day $t$ is drawn from a multinomial distribution:
\begin{equation} \label{eq:Zt}
    Z_{st} \sim \mathrm{Multinomial}(I_s, u_{t-s}), \qquad \qquad t-s=0,1,2,\ldots
\end{equation}
Note the infection-to-notification time encompasses the incubation period plus any additional delay from symptom onset to notification. This could be explicitly modelled using two separate distributions if required, but we here we consider a single distribution for the overall time from infection to notification.

The total number of individuals with a notification time on day $t$ is $\sum_{s=1}^t Z_{st}$. We assume that infections have a fixed probability $\alpha\in(0,1]$ of being notified as cases. {\highlight Hence, the expected number of case notifications on day $t$ is $\mu=\alpha \sum_{s=1}^t Z_{st}$.} To allow for noise in daily case notifications, we use a negative binomial distribution for the observed number of case notifications $C_t$ on day $t$ \cite{abbott2020estimating}:
\begin{equation} \label{eq:Ct}
    C_t \sim \mathrm{NegBin}\left(  \mu, k_c\right),     
\end{equation}
where $k_c$ is the dispersion parameter for the observed data (note that unlike $k_r$, $k_c$ does not impact the transmission process, only the variance of the observed data). Whilst we retain the negative binomial distribution for model flexibility, in practice we set $k_c=\infty$, meaning that the number of daily observed cases is Poisson distributed.

\subsection*{Fitting method}

We fit the model to data on daily local case notifications using a bootstrap particle filter \cite{gordon1993novel}, using data on imported cases (where available) as seed infections. We simulate a set of $m=10^5$ particles according to Eqs. \eqref{eq:Rt}--\eqref{eq:Zt}. {\highlight At each daily time step $t$, the likelihood $w^{(j)}$ for each particle $j=1,\ldots,m$ is calculated using the data $C_t^\mathrm{obs}$ for the number of case notifications on day $t$. It follows from Eq. \eqref{eq:Ct} that the likelihood is given by
\begin{equation} \label{eq:likelihood}
w^{(j)} = f_{NB}\left(C_t^\mathrm{obs};  \mu^{(j)}, k_c\right),
\end{equation}
where $\mu^{(j)}$ is the expected number of case notifications on day $t$ for particle $j$, and $f_{NB}$ is the probability mass function for a negative binomial distribution with mean $\mu^{(j)}$ and dispersion parameter $k_c$.}

Particles are then resampled, with replacement, with probability $w^{(j)}$. After resampling, the particles represent samples from the posterior distribution over the hidden states conditional on the observed data up to day $t$ \cite{sarkka2023bayesian}.
To avoid sample degeneracy in the time series of the particles, we used fixed-lag resampling, which means that instead of resampling the entire history of each particle at each time step, we only resample the most recent $t_\mathrm{lag}$ days \cite{watson2024jointly}.

\subsection*{End-of-outbreak probabilities}
As in \cite{thompson2024using}, the probability that there are are no new infections on or after day $t$, assuming a constant reproduction number $R$, is given by
\begin{equation} \label{eq:P0}
    P_0 = e^{-R \gamma_t},
\end{equation}
where 
\begin{equation} \label{eq:gammat}
    \gamma_t = \sum_{s=1}^{t-1} Y_{t-s} (1-F_{s-1}),
\end{equation}
and $F$ is the cumulative distribution function for the generation time distribution, defined by $F_s=\sum_{s'=1}^s g_{s'}$. Note that in Eq. \eqref{eq:gammat}, $I_{t-s}$ is replaced by $Y_{t-s}$ to account for the individual heterogeneity in transmission rates and the effect of imported infections. 

Because our model explicitly includes delays from infection to notification, it is possible that there could still be cases notified on or after day $t$, even if there are no new infections. To account for this, we also defined the probability $P_{00}$ that there will be no new infections and no notified cases on or after day $t$. As shown in Supplementary Material Section S1, $P_{00}$ may be expressed as
\begin{equation}  \label{eq:phit}
    P_{00} = e^{-R \gamma_t} \phi_t, \qquad \textrm{where } 
\phi_t = \prod_{t'=t}^\infty \left(\frac{k_c}{k_c+\alpha\sum_{s=1}^{t-1} Z_{st'}} \right)^{k_c}.
\end{equation}

We also calculated the probability of ultimate extinction (i.e. probability that the number of active infections eventually becomes zero, even with an unlimited susceptible population) under constant reproduction number $R$:
\begin{equation} \label{eq:PUE}
    \mathrm{PUE} = e^{-(1-q) R \gamma_t},
\end{equation}
where $q\in[0,1]$ is the probability of ultimate extinction for an outbreak that starts with a fully infectious seed case (see Supplementary Material Section S1 for full derivation).
PUE is the probability that the transmission chains stemming from any current infections self-extinguish, as opposed to generating a large outbreak that only ends due to depletion of the susceptible population (or reintroduction of control measures). This is equivalent to the ``probability of stochastic extinction'' defined by \cite{lloyd2005superspreading}, but adapted to consider starting with all currently active infections at time $t$.
In general, PUE will be larger than $P_0$ as there is a possibility that there will be one or more future infections but the outbreak nevertheless self-extinguishes due to stochastic effects. 

We calculated the values of $\gamma_t$ and $\phi_t$ for each of the $m$ particles according to Eqs. \eqref{eq:gammat} and \eqref{eq:phit} on each day $t$, before resampling the particles according to the likelihood function given the data on the number of case notifications on day $t$. This ensured that the end-of-outbreak probabilities were real-time estimates, i.e. are based only on data that was available prior to day $t$ and not subsequent data.

{\highlight The pre-intervention reproduction number, denoted $R_t^\mathrm{pre}$, for each of the $m$ particles was defined to be the average estimated value of $R_t$ in the 14 days prior to the start of the intervention. This value was calculated conditional on data up to the start of the intervention but not subsequent data. This provided a sample of $m$ values from the distributional estimate for $R_t^\mathrm{pre}$.}

{\highlight We defined $P_0$, $P_{00}$ and PUE as the probabilities of the respective outcome, under the assumption that the intervention was lifted on day $t$ and the reproduction number consequently reverted to its pre-intervention value. We estimated $P_0$, $P_{00}$ and PUE for each particle at each time $t$ by choosing a random sample from the distribution of $R_t^\mathrm{pre}$ and using this as the value of $R$ in Eqs. \eqref{eq:P0}, \eqref{eq:phit} and \eqref{eq:PUE}. The overall estimates for $P_0$, $P_{00}$ and PUE were found by averaging over all $m$ particles. These estimates therefore include the effect of uncertainty in the estimated pre-intervention reproduction number. Note all end-of-outbreak probability estimates assumed that there would be no further imported infections on or after day $t$. }

\subsection*{Case study 1: Covid-19, New Zealand, 2020}

For the New Zealand Covid-19 outbreak, we used Ministry of Health data on the daily number of Covid-19 case notifications in New Zealand between February and June 2020 \cite{nz_covid_data}. We assumed that the generation time (for wildtype SARS-CoV-2) was a gamma distribution with mean 5.05 days and standard deviation 1.94 days \cite{ferretti2020quantifying}. We assumed that the time from infection to notification was a gamma distribution with mean $\pm$ standard deviation of either 7.7 $\pm$ 3.2 days or 11.2 $\pm$ 4.7 days. These correspond to  an incubation period of 5.5 $\pm$ 2.3 days, plus values reported by Hendy et al. \cite{hendy2021mathematical} for the self-reported delay from symptom onset to isolation and from symptom onset to notification, respectively. We use these as indicative values representing a short and a long notification delay scenario respectively, and note that exact values will be outbreak-specific. We truncated the generation time and reporting time distributions at a maximum of 15 days and 25 days respectively, and set $t_\mathrm{lag}=30$ days to ensure the resampling lag was greater than both of these maxima \cite{watson2024jointly}. 

Cases were classified by the Ministry of Health as either local or imported.
We assumed that the date of infection for all imported cases was $5$ days prior to date of notification (which means that the first imported case had an infection date of 21 February 2020). We also assumed that imported infections had the same notification probability $\alpha$ as local infections.  To model this, we generated a set of unreported imported infections with infection dates resampled with replacement from the infection dates for the notified imported cases. We assumed that imported infections had an average infectivity of $\varepsilon=0.5$ relative to local cases, representing the effect of self-isolation measures for arriving travellers. This value for $\varepsilon$ is consistent with the findings of \cite{james2021model} based on contact tracing data. For simplicity we assumed that imported and local cases had the same generation time distribution. We assumed that any imported cases with an infection date after the introduction of mandatory government-managed quarantine (10 April 2020) had a negligible risk of causing local infections and could therefore be ignored. 

Because the case ascertainment rate was highly uncertain, we tested different values for the notification probability $\alpha$ of $0.4$ and $0.7$. We also ran the model for $\alpha=1$ to provide a baseline for assessing the impact of including under-reporting in the model. We also investigated different values for the transmission overdispersion parameter of $k_r=\infty$ (representing a Poisson offspring distribution, i.e. no heterogeneity in transmission multipliers), $k_r=1$ and $k_r=0.2$ (representing increasing levels of heterogeneity). These values span the estimated value of $k_r=0.41$ for SARS-CoV-2 from a recent meta-analysis \cite{wegehaupt2023superspreading}. 

To initialise the model, we assumed that there were zero infections prior to the first imported infection. We drew the value of $R_t$ at the start of the simulation from a prior distribution, assumed to be a Gamma distribution with mean $2$ and standard deviation $1$. We set the time window for the rapid change in $R_t$ to be 7 days starting on 23 March 2020, the day on which it was announced that the country would move to ``Alert Level 4'' two days subsequently. During this $7$-day time window, we set the daily random walk step to have mean $\delta_t=-0.1$ and standard deviation $\sigma_t=0.2$. {\highlight  This corresponds to a broad prior for the effect of the intervention on $R_t$, with an expected aggregate reduction in $R_t$ by a factor of $\exp(-7\times 0.1)\approx 0.5$, which is approximately consistent with previous analyses \cite{hendy2021mathematical}. The chosen values were also found visually to give a good qualitative fit to the data and we provide a sensitivity analysis in Supplementary Material Section S2).} All model parameter values are shown in Table \ref{tab:params}.

\subsection*{Case study 2: EVD, \'Equateur Province, DRC, 2018}

We also analysed data from an EVD outbreak with 54 cases that occurred in \'Equateur Province of DRC in 2018. Following laboratory confirmation of the first two cases on 8 May, the start of the outbreak was declared and a response team (the Ebola Response Team; ERT) was deployed by the DRC government and international partners \cite{nkengasong2018response,mbala20192018}. The ERT implemented a range of measures, including ring vaccination, active case finding, contact tracing, case isolation and treatment, laboratory testing, and community engagement. On 24 July, 42 days after the final case recovered, the outbreak was declared over and the ERT was withdrawn \cite{mbala20192018}.

Symptom onset dates were available for all cases so, consistent with previous analyses of this outbreak \cite{thompson2024using}, we analysed the data by date of symptom onset rather than date of notification. This is unlikely to have a substantial effect on model results because, in the latter stages of the outbreak when end-of-outbreak probabilities become relevant, the notification date was the same as the symptom onset date for most cases.  

We assumed the generation time was a gamma distribution with mean 15.3 days and standard deviation 9.3 days. This corresponds to the serial interval estimate of \cite{vankerkhove2015review}. However, since the generation time and serial interval distributions for EVD are thought be very similar, we contend that it is reasonable to use this as the generation time distribution \cite{who2014ebola}. We tested two different values for the mean and standard deviation of the infection-to-notification time distribution (see Table \ref{tab:params}). These correspond to two different estimates for the mean and standard deviation of the incubation period for EVD  \cite{eichner2011incubation,velasquez2015time}. We truncated the generation time and reporting time distributions at a maximum of 50 days and 30 days respectively, and  set $t_\mathrm{lag}=50$ days. We ran the model for three different values for the notification probability $\alpha$ of 0.8, 0.9 and 1 (we considered these relatively high values due to the extensive case finding activities undertaken by the ERT).  We also investigated the same three values for the overdispersion parameter $k_r$ as for the Covid-19 outbreak. These values span the value of $k_r=0.18$ estimated for EVD by Althaus \cite{althaus2015ebola}

To initialise the model, we assumed that the first case resulted from a single zoonotic spillover event and all subsequent cases resulted from local human-to-human transmission. We drew the value of $R_t$ at the start of the simulation from a prior distribution, assumed to be a Gamma distribution with mean $2.5$ and standard deviation $1$. We set the time window for the rapid change in $R_t$ to be 7 days starting on 8 May 2018, the date on which the ERT arrived. During this time window, we set the daily random walk step to have mean $\delta_t=-0.1$ and standard deviation $\sigma_t=0.2$, as for the Covid-19 outbreak model.

\begin{table}
    \centering
    \begin{tabular}{p{9cm}ll}
    \hline
    Parameter & {\bf Covid-19} & {\bf EVD} \\
    \hline

      Reproduction number prior mean & 2.0 & 2.5 \\
      Reproduction number prior s.d. & 1.0 & 1.0 \\
      Generation time mean & 5.05 days & 15.3 days \\
      Generation time s.d. & 1.94 days & 9.3 days \\
      Time from infection to notification mean  & [{\bf 7.7}, 11.2] days & [{\bf 6.2}, 11.2] days \\
      Time from infection to notification s.d. & [{\bf 3.2}, 4.7] days & [{\bf 1.6}, 4.3] days  \\
      Probability of notification ($\alpha$) & [{\bf 0.4}, 0.7, 1] & [{\bf 0.8}, 0.9, 1]  \\
      Dispersion parameter for individual transmissibility ($k_r$) & [$0.2$, $1$, $\boldsymbol{\infty}$] & [$0.2$, $1$, $\boldsymbol{\infty}$] \\
      Dispersion parameter for daily case notifications ($k_c$) & $\infty$ & $\infty$ \\
      Relative transmission potential for imported cases ($\varepsilon$) & 0.5 & 1 \\
      Start of intervention-related change in $R_t$ & 23 Mar 2020 & 8 May 2018  \\
      Duration of intervention-related change in $R_t$ & 7 days & 7 days \\
      Reproduction number random walk step mean during change window ($\delta_c$) & $-0.1$ & $-0.1$ \\   
      Reproduction number random walk step s.d. during change window ($\sigma_c$) & $0.2$ & $0.2$ \\
      Reproduction number random walk step s.d. at other times ($\sigma_R$) & $0.05$ & 0.05 \\
      Particle filter resampling lag ($t_\mathrm{lag}$) & 30 days & 50 days \\
    \hline
    \end{tabular}
    \caption{Parameter values used to model the Covid-19 outbreak in New Zealand in 2020 and the EVD outbreak in DRC in 2018. The reproduction number prior distribution, generation time distribution, and infection-to-notification time distributions were gamma distributions with the specified mean and standard deviation (s.d.). Where multiple parameter values were investigated, the default value is indicated in bold. }
    \label{tab:params}
\end{table}

\section*{Results}

\subsection*{Covid-19, New Zealand, 2020}

\begin{figure}
    \centering
    \includegraphics[width=\linewidth]{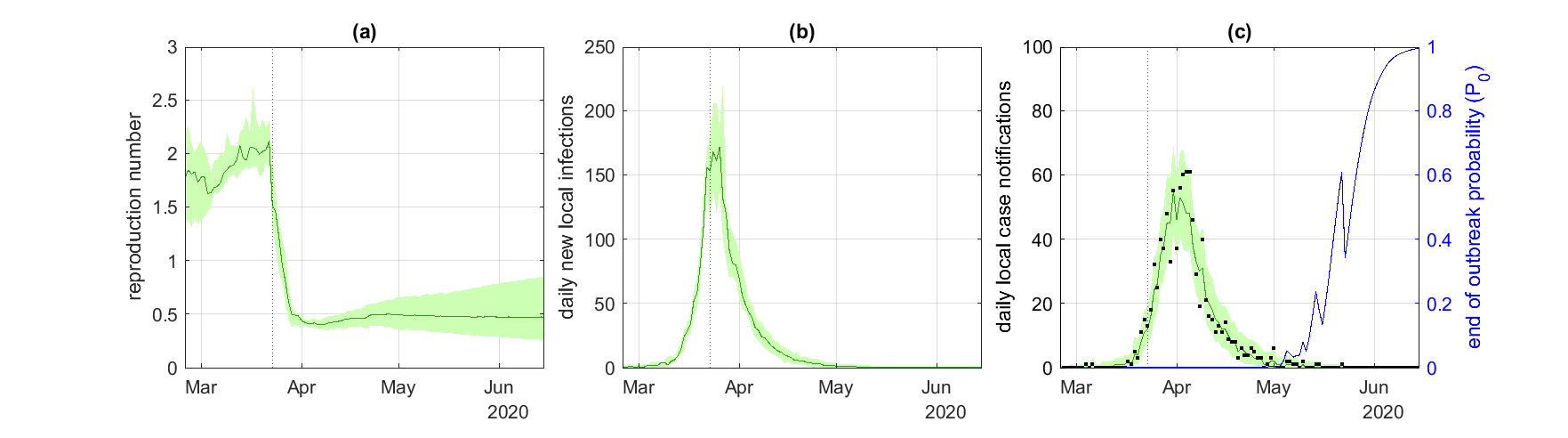}
    \caption{Model results for the New Zealand Covid-19 outbreak: (a) instantaneous reproduction number $R_t$; (b) daily new local infections $I_t$; (c) daily local case notifications $C_t$ and probability $P_0$ of no future infections. Green curve shows the median and shaded band shows the 5th and 95th percentiles of $m=10^5$ particles; black points show data for daily local case notifications. Dotted vertical line shows the start of the 7-day time window in which a rapid drop in the reproduction number was expected {\em a priori}. Notification probability $\alpha=0.4$, mean time from infection to notification $t_n=7.7$ days, dispersion parameter $k_r=\infty$ (i.e. offspring distribution was Poisson).}
    \label{fig:1}
\end{figure}

The daily number of locally acquired Covid-19 case notifications in New Zealand peaked at 61 on 4 April 2020 and the last notified case of the outbreak was on 22 May 2020. For the default model parameters (see Table \ref{tab:params}), the median reproduction number varied between $1.7$ and $2.2$ in the period prior the start of the intervention, falling to between $0.4$ and $0.5$ after the intervention on 25 March 2020 (Figure \ref{fig:1}a). With the assumed case ascertainment rate of $\alpha=0.4$, new local infections peaked at approximately 150--200 per day (Figure \ref{fig:1}b) and the model provided a good qualitative fit to the data on local case notifications (Figure \ref{fig:1}c). 

The estimated probability $P_0$ that there would be no future infections if the reproduction number reverted to its pre-intervention value first became non-zero around the beginning of May 2020 (Figure \ref{fig:1}c, blue curve). The value of $P_0$ tended to increase gradually during intervals with no new case notifications, and to drop sharply when sporadic new cases were notified, for example on 14-15 May and 22 May. After the last case notification on 22 May, $P_0$ increased steadily from around 30\%, reaching 95\% on 6 June 2020 (hereafter, when describing all of our analyses in the text, we refer to relevant probabilities as percentage values). The actual date on which the outbreak was declared to have been eliminated and control measures were relaxed was 8 June 2020. {\highlight Note a threshold value of 95\% for the end-of-outbreak probability is chosen here for illustrative purposes only. Other choices of threshold are equally possible and in reality the appropriate choice of threshold for relaxation of control measures will depend on the epidemiological and socioeconomic context (see Discussion for more detail).} {\highlight See Supplementary Material Section S2 for additional discussion regarding uncertainty in the estimated value of $P_0$.}

\begin{figure}
    \centering
    \includegraphics[width=\linewidth]{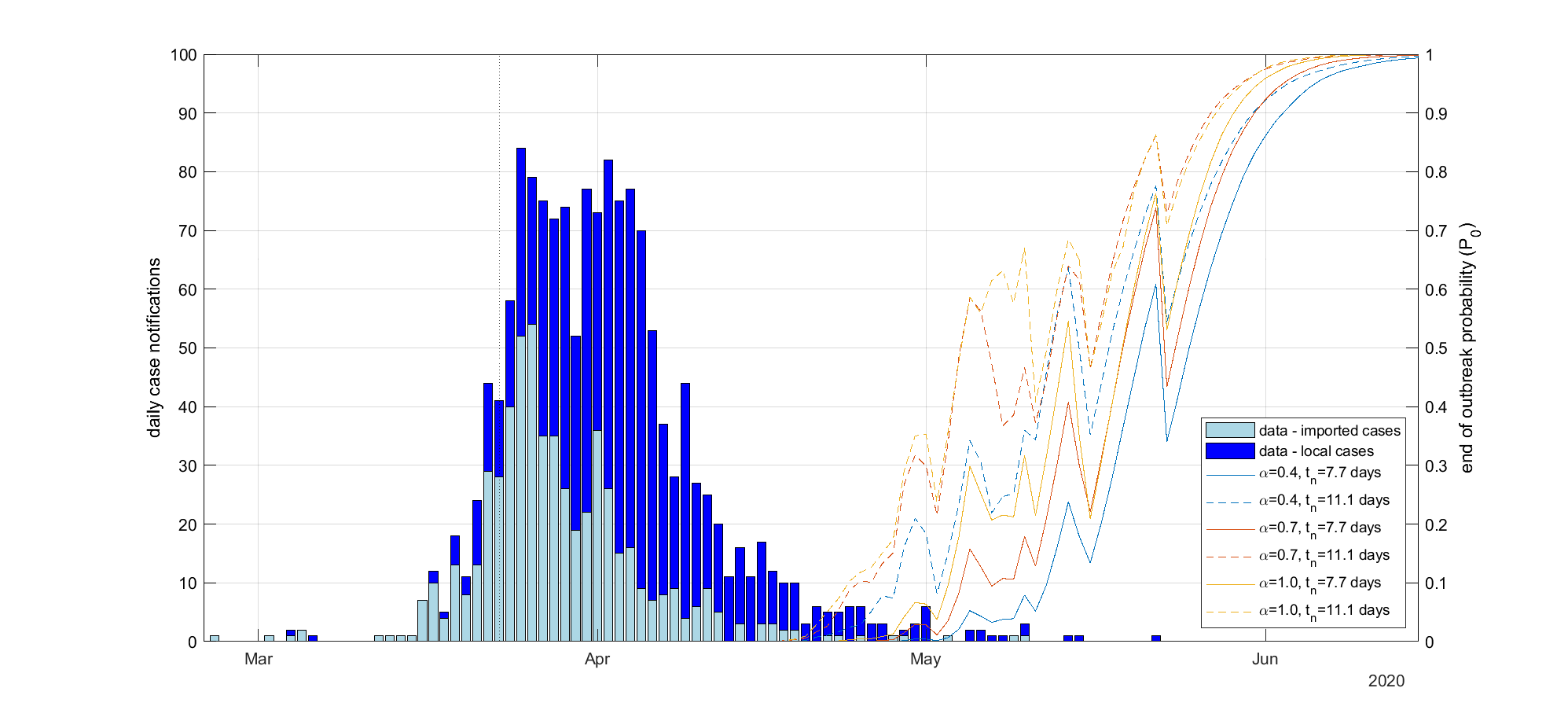}
    \caption{Daily number of imported and local case notifications (bars) and the probability $P_0$ of no future infections (curves) for the New Zealand Covid-19 outbreak with different values for the notification probability $\alpha$ and mean infection-to-notification time $t_n$. Dispersion parameter $k_r=\infty$ (i.e. offspring distribution was Poisson).}
    \label{fig:2}
\end{figure}

\begin{figure}
    \centering
    \includegraphics[width=\linewidth]{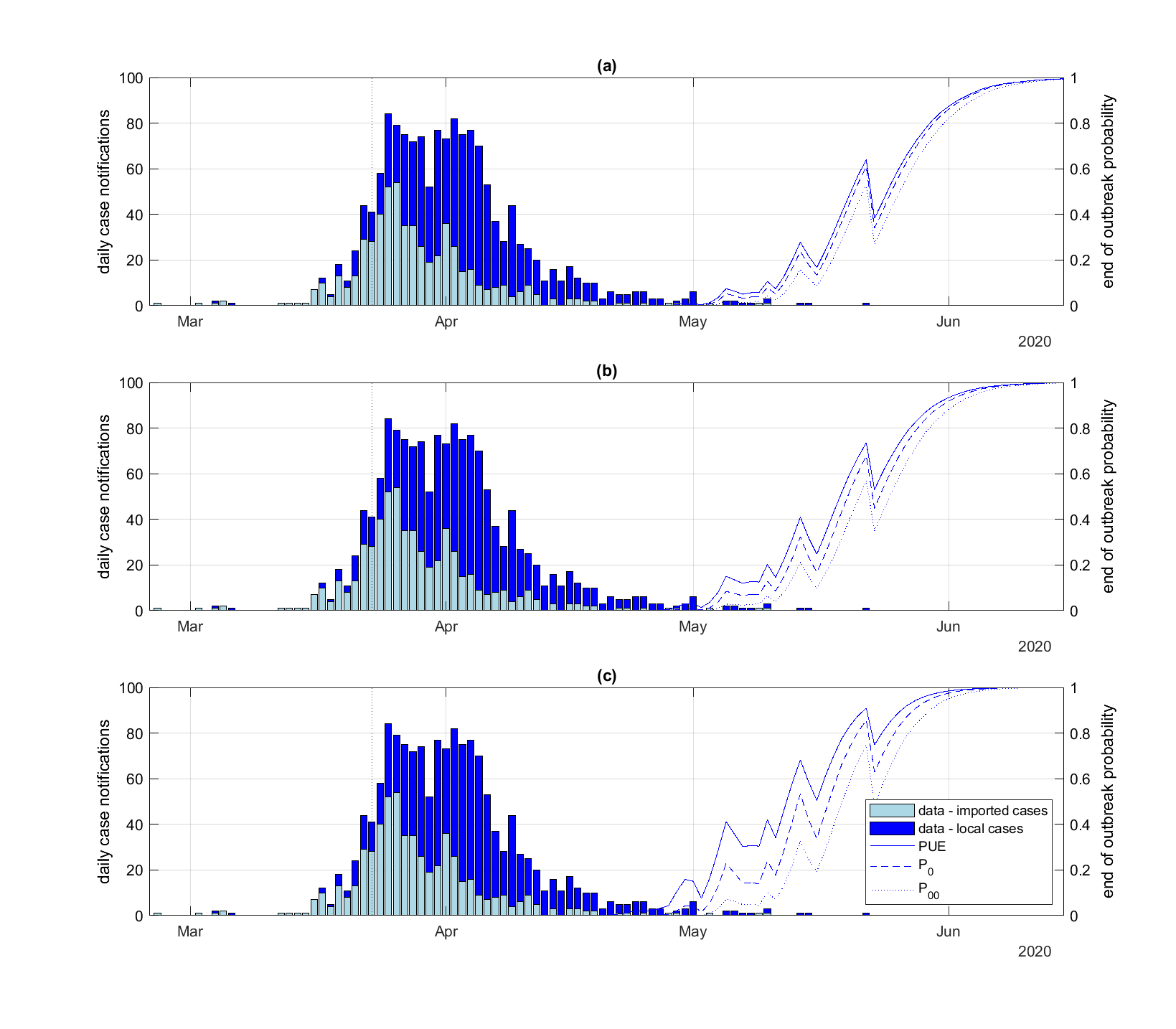}
    \caption{Daily number of imported and local case notifications (bars) alongside end-of-outbreak probabilities (curves) for the New Zealand Covid-19 outbreak with different offspring distribution dispersion parameters: (a) $k_r=\infty$ (i.e. Poisson offspring distribution; (b) $k_r=1$; (c) $k_r=0.2$. Each panel shows the probability of ultimate extinction (solid), the probability $P_0$ of no future infections (dashed) and the probability $P_{00}$ of no future infections or notifications (dotted). Notification probability $\alpha=0.4$, mean time from infection to notification $t_n=7.7$ days.}
    \label{fig:3}
\end{figure}

\begin{table}[]
    \centering
\begin{tabular}{lll} 
\hline
\multicolumn{3}{l}{ \bf Covid-19} \\ 
 &  $t_n=7.7$ days & $t_n=11.2$ days \\ 
$\alpha=0.4$ & 06-Jun-2020 & 04-Jun-2020 \\ 
 $\alpha=0.7$ & 03-Jun-2020 & 30-May-2020 \\ 
 $\alpha=1.0$ & 01-Jun-2020 & 30-May-2020 \\ 
 \hline
\multicolumn{3}{l}{ \bf Ebola} \\ 
 &  $t_n=6.2$ days & $t_n=11.2$ days \\ 
$\alpha=0.8$ & 30-Jul-2018 & 25-Jul-2018 \\ 
 $\alpha=0.9$ & 30-Jul-2018 & 22-Jul-2018 \\ 
 $\alpha=1.0$ & 26-Jul-2018 & 22-Jul-2018 \\ 
 \hline
\end{tabular} 

\caption{Date on which the estimated probability $P_0$ of no future infections first exceeded 95\% {\highlight (note this is an arbitrary threshold chosen for illustrative purposes only)}.   Results are shown for the New Zealand Covid-19 outbreak and DRC EVD outbreak for low, medium and high notification probability $\alpha$ and for short and long mean infection-to-notification time $t_n$. Dispersion parameter $k_r=\infty$ (i.e. offspring distribution was Poisson). }
    \label{tab:elimination_dates}
\end{table}

Increasing the assumed notification probability $\alpha$ led to higher estimated values of $P_0$ (Figure \ref{fig:2}, solid red and yellow curves). For example, the value of $P_0$ reached 95\% on 3 June 2020 for $\alpha=0.7$ and on 1 June 2020 for $\alpha=1$ (Table \ref{tab:elimination_dates}). This was because a higher assumed case ascertainment rate meant that, for a given number of case notifications, the inferred number of infections was lower and  therefore reached zero sooner. However, it was notable that the difference between the estimated time at which $P_0$ reached 95\% for $\alpha=0.4$ and $\alpha=1$ was only 5 days. Since end-of-outbreak declarations based on this type of analysis would only be expected to be made when the estimated end-of-outbreak probability is close to 1, this result suggests that end-of-outbreak declaration dates may in some situations be relatively insensitive to the assumed case ascertainment rate.

Increasing the mean time $t_n$ from infection to notification from $7.7$ days to $11.2$ days meant that the estimated value for $P_0$ started to increase earlier and generally was higher at any given time than with a smaller $t_n$ (Figure \ref{fig:2}, dashed curves). This counterintuitive result occurred because, during the intervention period, the estimated reproduction number was well under the critical value of $1$, meaning that incidence of new infections was on an exponentially decaying trajectory. Observed data at time $t$ gives information about infections that happened around approximately $t-t_n$. Hence, the larger $t_n$, the longer incidence has been exponentially decaying in the period between infection and observation. Notably however, the time at which $P_0$ reached 95\% was only $2$--$4$ days earlier with $t_n=11.2$ days than with $t_n=7.7$ days (Table \ref{tab:elimination_dates}). {\highlight Additional results showing the hidden states $R_t$ and $I_t$ for different values of $\alpha$ and $t_n$ are provided in Supplementary Figure S1.}

We also investigated model results for different levels of individual heterogeneity in transmission and using different definitions for the end-of-outbreak probability. The probability $P_{00}$ of no future infections or notifications (Figure \ref{fig:3}a, dotted curves) was lower than the probability of no future infections ($P_0$). {\highlight This is because the delay from infection to notification means that it is possible for cases to be reported on or after day $t$ even if there are no new infections.} In contrast, the probability of ultimate extinction (PUE) (Figure \ref{fig:3}a, solid curves) was higher than $P_0$. {\highlight This is because transmission chains have a non-zero probability of self-extinguishing, even when the reproduction number is greater than $1$.}

Increasing the level of individual heterogeneity in transmission (i.e. reducing the value of the dispersion parameter $k_r$) led to higher estimates for the end-of-outbreak probability, whichever of three definitions was used (Figure \ref{fig:3}b,c). This was expected as it is well known that increasing individual heterogeneity in the number of secondary infections (decreasing $k_r$) increases the likelihood of stochastic extinction \cite{lloyd2005superspreading}. For example, with $k_r=0.2$ (Figure \ref{fig:3}c), the probability  $P_0$ of no future infections reached 95\% on 30 May 2020, which was 7 days earlier than in the scenario without heterogeneity (Figure \ref{fig:3}a). {\highlight Additional results showing the hidden states $R_t$ and $I_t$ for different values of $k_r$ are provided in Supplementary Figure S2.

A sensitivity analysis on the random walk parameters for the time-varying reproduction number is provided in Supplementary Figure S3 and shows that the results are not highly sensitive to these parameters.}

\subsection*{EVD, \'Equateur Province, DRC, 2018}
\begin{figure}
    \centering
    \includegraphics[width=\linewidth]{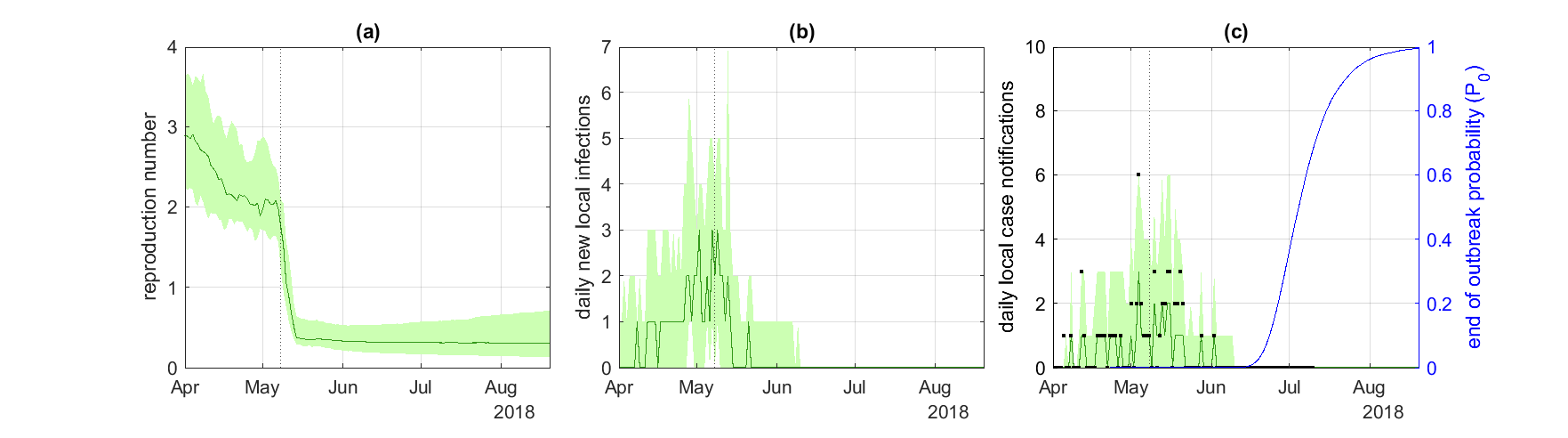}
    \caption{Model results for the DRC EVD outbreak: (a) instantaneous reproduction number $R_t$; (b) daily new local infections $I_t$; (c) daily local case notifications $C_t$ and probability $P_0$ of no future infections. Green curve shows the median and shaded band shows the 5th and 95th percentiles of $m=10^5$ particles; black points show data for daily local case notifications. Dotted vertical line shows the start of the 7-day time window in which a rapid drop in the reproduction number was expected {\em a priori}. Notification probability $\alpha=0.8$, mean time from infection to notification $t_n=6.2$ days, dispersion parameter $k_r=\infty$ (i.e. offspring distribution was Poisson).}
    \label{fig:4}
\end{figure}

We went on to analyse the 2018 \'Equateur Province EVD outbreak dataset. The highest number of daily cases was six on 4 May 2018, just before the arrival of the ERT on 8 May. For the default model parameters (see Table \ref{tab:params}), the median reproduction number varied between $2.0$ and $2.9$ in the period prior to the start of the intervention, falling to approximately $0.3$--$0.4$ after the intervention (Figure \ref{fig:4}a). 

Because of the longer generation time of EVD compared to Covid-19, the estimated probability $P_0$ of no future infections did not start to increase substantially above zero until around 2 weeks after the last reported case on 2 June 2018, and subsequently increased smoothly towards $1$ (Figure \ref{fig:4}c). This contrasted with the results for Covid-19 where the estimated end-of-outbreak probabilities started to increase after a few days with no newly reported cases but dropped down after a new case notification, leading to sawtooth-shaped curves. 

\begin{figure}
    \centering
    \includegraphics[width=\linewidth]{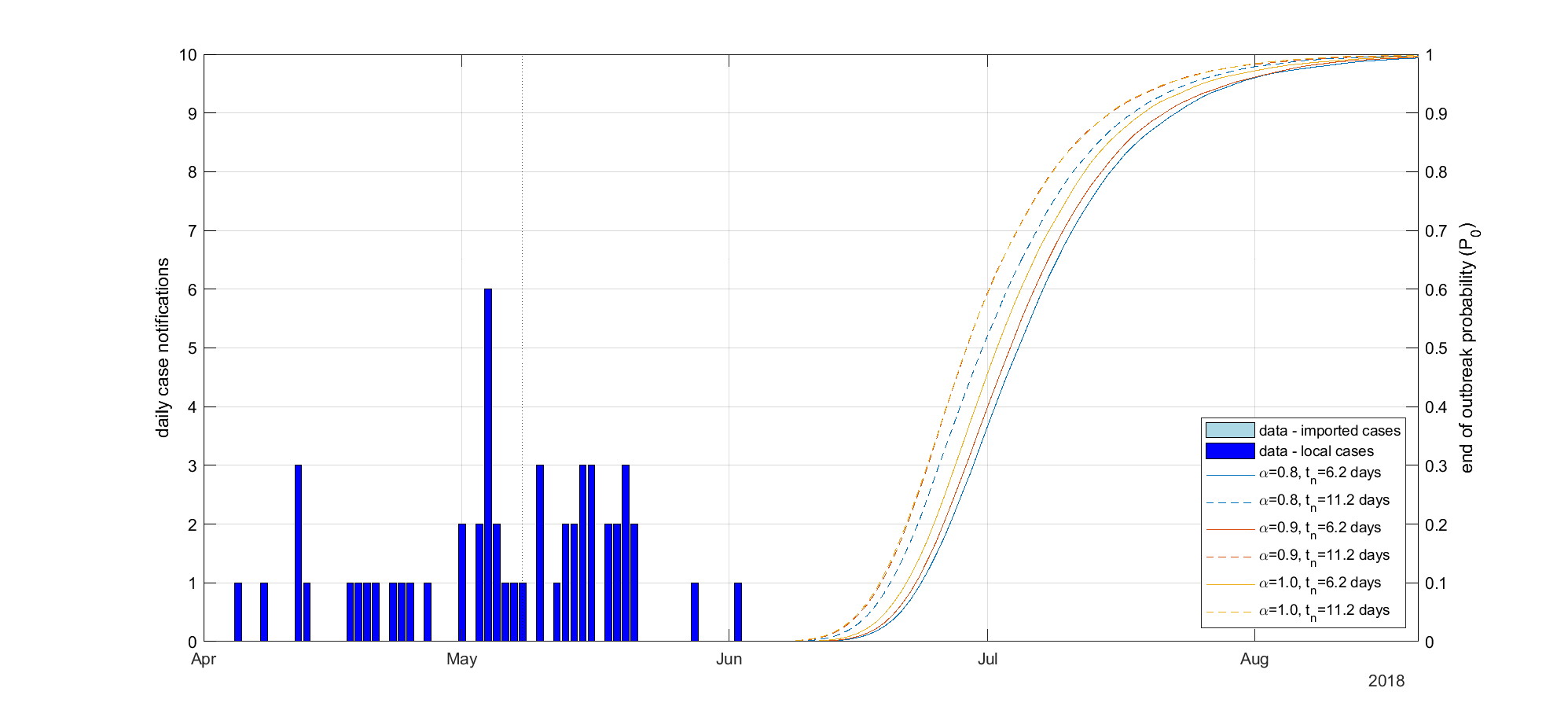}
    \caption{Daily number of case notifications (bars) and probability $P_0$ of no future infections (curves) for the DRC EVD outbreak with different values for the notification probability $\alpha$ and mean infection-to-notification time $t_n$. Dispersion parameter $k_r=\infty$ (i.e. offspring distribution was Poisson).}
    \label{fig:5}
\end{figure}

The effects of the notification probability $\alpha$ and mean time to notification $t_n$ on the estimated probability $P_0$ of no future infections were broadly similar for EVD as for Covid-19. Increasing either $\alpha$ or $t_n$ tended to increase $P_0$ (Figure \ref{fig:5}). For the parameter values investigated, the estimated probability $P_0$ of no future infections reached 95\% in the range 22 July to 30 July 2018 (Table \ref{tab:elimination_dates}). The actual date on which the ERT was withdrawn was 24 July 2018. {\highlight Additional results showing the hidden states $R_t$ and $I_t$ for different values of $\alpha$ and $t_n$ are provided in Supplementary Figure S4.}

\begin{figure}
    \centering
    \includegraphics[width=\linewidth]{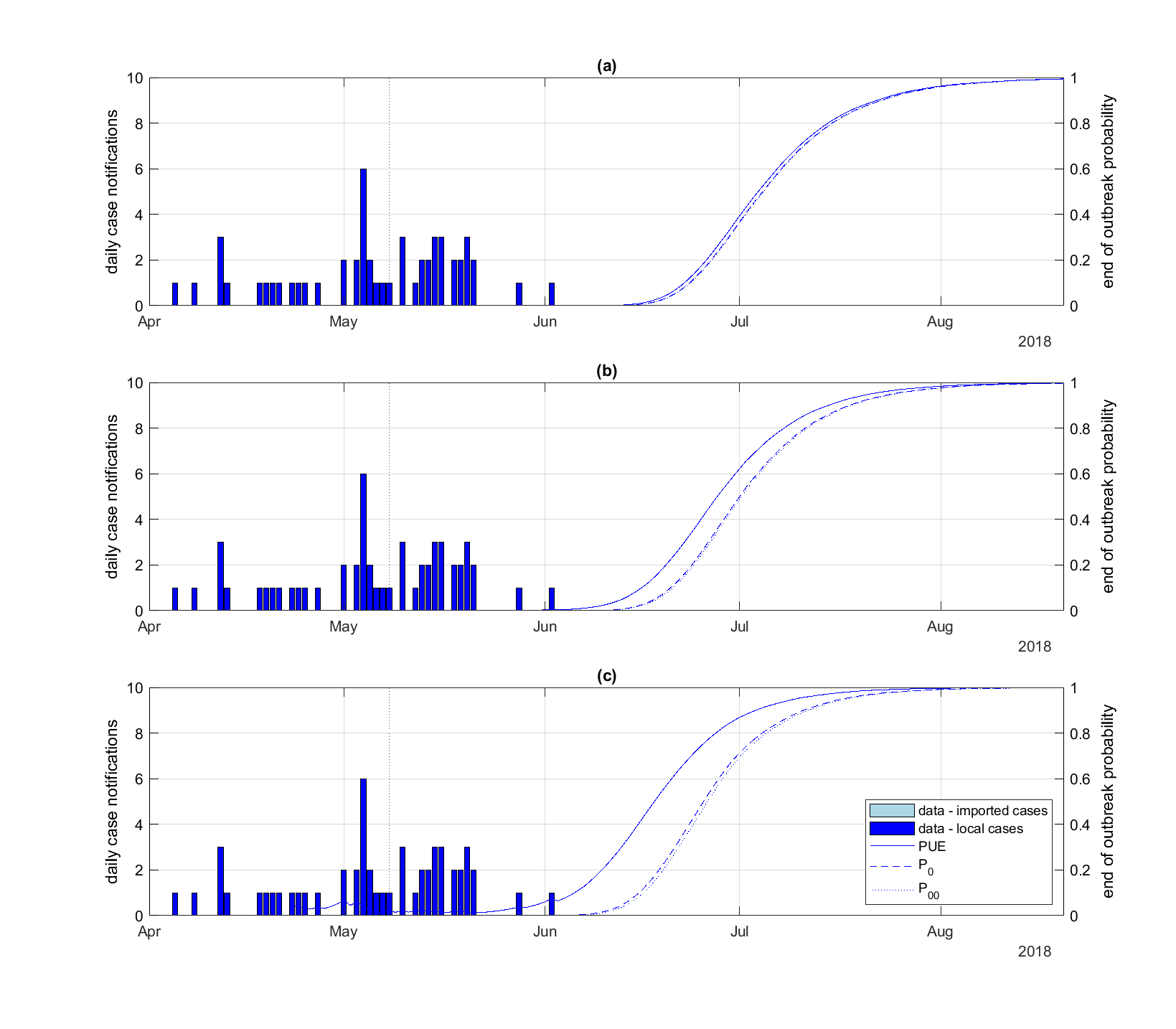}
    \caption{Daily number of imported and local case notifications (bars) alongside end-of-outbreak probabilities (curves) for the DRC EVD outbreak with different offspring distribution dispersion parameters: (a) $k_r=\infty$ (i.e. Poisson offspring distribution; (b) $k_r=1$; (c) $k_r=0.2$. Each panel shows the probability of ultimate extinction (solid), the probability $P_0$ of no future infections (dashed) and the probability $P_{00}$ of no future infections or notifications (dotted). Notification probability $\alpha=0.8$, mean time from infection to notification $t_n=6.2$ days }

    \label{fig:6}
\end{figure}

The relative behaviour of the alternative definitions of the end-of-outbreak probability and the effect of heterogeneous transmission patterns were also similar for EVD as for Covid-19. Reducing the dispersion parameter $k_r$ increased the estimated end-of-outbreak probabilities (Figure \ref{fig:6}). This was particularly true for the PUE (Figure \ref{fig:6}, solid curves), which, for the most highly overdispersed scenario (Figure \ref{fig:6}c) was approximately $0.1$ even during the middle of the outbreak. This reflects the fact that small outbreaks with highly overdispersed transmission patterns have a significant probability of stochastic extinction, even if the reproduction number is well above $1$. 

The other two end-of-outbreak metrics, $P_0$ and $P_{00}$ (Figure \ref{fig:6}, dashed and dotted curves respectively), were very similar to each other for EVD. This was a consequence of the fact that the generation time was long relative to the time from infection to notification, meaning the likelihood of there being cases yet to be notified but no longer infectious was small.  {\highlight Additional results showing the hidden states $R_t$ and $I_t$ for different values of $k_r$ are provided in Supplementary Figure S5. A sensitivity analysis on the random walk parameters for the time-varying reproduction number is provided in Supplementary Figure S6 and shows that the results are not highly sensitive to these parameters.}

\section*{Discussion}
At the tail end of an infectious disease outbreak, a key question for public health policy makers is when the outbreak can be declared over safely, signalling that interventions can be relaxed or removed \cite{nishiura2016methods, linton2022epidemics}. This is particularly important and challenging for diseases that have a high health or economic impact, but with a significant fraction of asymptomatic or subclinical infections and/or long incubation or infectious periods \cite{klepac2015six}. 

We have developed a flexible mathematical framework for estimating three versions of the end-of-outbreak probability, specifically: (i) the probability of no future infections ($P_0$); (ii) the probability of no future infections or notifications from already active infections ($P_{00}$); and (iii) the probability of ultimate extinction (PUE; i.e., the probability that already active infections do not generate sustained future chains of transmission). We calculated each of these quantities under the assumption that the reproduction number would return to its pre-intervention value once interventions are removed. 

Inference of $P_0$, $P_{00}$ and PUE has the capability to support public health officials to make evidence-based decisions that balance the risk of outbreak resurgence with the costs of continued intervention. The main benefit of our framework relative to previous work in this area is that it simultaneously incorporates the effects of lagged and incomplete case reporting, individual heterogeneity in the number of secondary infections, and the source location of cases (when relevant data are available). 

We have applied our framework to case studies from two very different infectious disease outbreaks, showcasing its adaptability to different epidemiological contexts.  

Contemporaneous modelling of the 2020 Covid-19 outbreak in New Zealand estimated that the probability of no future infections, $P_0$, reached 95\% on 1 June under optimistic assumptions about case ascertainment, and on 13 June under pessimistic assumptions \cite{hendy2021mathematical}. The government formally declared that the outbreak had been eliminated and relaxed public health interventions on 8 June 2020. Our model estimated that the probability $P_0$ of no future infections reached 95\% on 30 May 2020 if case ascertainment was 70\% and the mean time to notification was $11.2$ days, and on 6 June 2020 if case ascertainment was 40\% and mean time to notification was $7.7$ days. Our framework has two main advantages over the individual-based model of \cite{hendy2021mathematical}: (i) it is a simpler model, requiring fewer modelling assumptions and parameter estimates; (ii) it updates end-of-outbreak probability estimates in real-time as new data become available, rather than being solely defined in terms of the number of consecutive days with no new case notifications.

For the 2018 EVD outbreak in DRC, the estimated probability of no future infections first reached 95\% on 22--30 July, depending on the assumed reporting rate and incubation period. This range of dates includes the actual date on which the outbreak was declared over and the ERT withdrawn (24 July, 42 days after the final case recovered). In comparison, when Thompson et al. \cite{thompson2024using} considered the same outbreak dataset, the end-of-outbreak probability reached 95\% in the range 9--21 July under different assumptions regarding the number of unreported cases (also using a different serial interval distribution to the one assumed here). We emphasise that a key advance of our study over previous work is that, {\highlight for a given reporting probability, our approach rigorously accounts for uncertainty in the number and timing of unreported cases, whereas Thompson et al. \cite{thompson2024using} relied on an approximate method for this. Specifically, in Supplementary Analysis 3 of the article by Thompson et al. \cite{thompson2024using}, each unreported case was added to the outbreak dataset on a random date that was sampled based on the force of infection each day generated by previously reported cases. This implicitly assumed that the infectors of unreported cases were reported, which in reality may not always be true. Our approach allows for the possibility that unreported cases were infected by either reported or unreported cases. }

As with all modelling studies, generating our results involved a range of assumptions. When considering theoretical end-of-outbreak declaration dates, we assumed that an outbreak would be declared over when the end-of-outbreak probability first exceeds 95\%. In general, the choice of threshold is flexible and should be determined by decision-makers depending on the public health context. Key factors in this decision include the costs associated with continuing the intervention and the impacts if additional cases occur. By providing alternative probability estimates with which to determine the end of an outbreak, we enable decision-makers to choose a more risk-averse or a more risk-tolerant approach depending on the context.

Our analyses highlight that end-of-outbreak probability estimates can vary based on different model inputs, including the case ascertainment rate, reporting delays and heterogeneity in transmission. These quantities are likely to vary even between different outbreaks of the same disease, and therefore estimates specific to the outbreak under consideration should be used if available. However, this may not always be possible, in which case a range of assumptions can be considered (as we did here). Notably, we found that in the two case studies we considered, model-informed end-of-outbreak declaration dates were relatively insensitive to assumed values for the case ascertainment rate and the mean time from infection to notification, under the assumption that an outbreak would only be declared over when the estimated end-of-outbreak probability took a low value. This suggests that our modelling framework remains useful even in the presence of uncertainty in parameter values. 

Including individual heterogeneity in transmission via an overdispersed offspring distribution increased estimated end-of-outbreak probabilities, consistent with previous literature \cite{lloyd2005superspreading}. In the case studies investigated, this meant that model-informed end-of-outbreak declaration dates occurred 1--2 weeks earlier than in the scenario with no heterogeneity. However, it should be noted that heterogeneity means that, if the outbreak does not go extinct, there is a greater risk that the number of infections could grow very rapidly \cite{james2007event}. Thus, the downside risks of relaxing public health interventions too early in the presence of heterogeneity are higher, and so a decision-maker may want to choose a more conservative threshold for the end-of-outbreak probability in this situation.

A key limitation of our analysis is that the case ascertainment rate is assumed to take a constant value, which is known at least approximately. In reality, the case ascertainment rate is likely to vary during an outbreak \cite{dalziel2018unreported} and requires estimation \cite{jarvis2022measuring}. However, we contend that variation in the case ascertainment rate may be lower in the latter stages of an outbreak (as considered here) than in the early stages when public awareness of the outbreak is changing. A range of methods exist for inference of case ascertainment rates during outbreaks \cite{gibbons2014measuring}, including undertaking community-based studies to measure case ascertainment precisely in a specific location and extrapolate this to the wider population. If the case ascertainment rate is uncertain, but is estimated to lie within a particular interval, our model can be run with values at the lower and upper end of the interval to provide an estimated range for the end-of-outbreak probability.

Finally, additional epidemiological complexity could be added to the transmission model underlying the results presented here. As an example, at the end of the 2014-2016 Ebola epidemic in west Africa, it was suggested that sexual transmission of EVD can occur several months after an infected individual recovers \cite{abbate2016potential}. If required, the potential for EVD recrudescence in this way can be built into the renewal equation model and resulting estimates of the end-of-outbreak probability \cite{lee2019sexual}. Isolation measures aimed at reducing transmission from confirmed cases could similarly be included in the model. 

Despite the assumptions underlying our research, we have provided a quantitative framework for estimating the end-of-outbreak probability that accounts for case underascertainment, reporting lags, heterogeneity in transmissibility between infectious individuals, and the source location of cases (i.e., imported or infected locally). Transmission dynamics near the end of an infectious disease outbreak tend, by their nature, to be highly stochastic and influenced by multiple factors. Public health decision-making needs to take account of the local context and circumstances of individual cases where relevant information is available. Our model framework is not intended as a substitute for human decision-making, but rather to provide a complementary source of quantitative evidence that can be weighed alongside qualitative epidemiological information and other sources of evidence as appropriate when deciding whether or not to declare an outbreak over.


\subsection*{Acknowledgements}
The authors would like to thank the Isaac Newton Institute for Mathematical Sciences, Cambridge, for support and hospitality during the programme ``Modelling and inference for pandemic preparedness'', where initial work on this research article was undertaken. This research was supported by EPSRC grant EP/R014604/1 (MJP, WSH and RNT). MJP acknowledges funding from Te Niwha Infectious Diseases Research Platform, Institute of Environmental Science and Research, grant number TN/P/24/UoC/MP. WSH and RNT would like to thank members of the Infectious Disease Modelling group in the Wolfson Centre for Mathematical Biology at the University of Oxford for useful discussions about this research. The authors are grateful to two anonymous reviewers for helpful comments.



\begin{thebibliography}{10}

\bibitem{tildesley2022optimal}
Tildesley MJ, Vassall A, Riley S, Jit M, Sandmann F, Hill EM, et~al.
\newblock Optimal health and economic impact of non-pharmaceutical intervention measures prior and post vaccination in England: a mathematical modelling study.
\newblock Royal Society Open Science. 2022;9(8):211746.

\bibitem{dobson2023balancing}
Dobson A, Ricci C, Boucekkine R, Gozzi F, Fabbri G, Loch-Temzelides T, et~al.
\newblock Balancing economic and epidemiological interventions in the early stages of pathogen emergence.
\newblock Science Advances. 2023;9(21):eade6169.

\bibitem{klepac2015six}
Klepac P, Funk S, Hollingsworth TD, Metcalf CJE, Hampson K.
\newblock Six challenges in the eradication of infectious diseases.
\newblock Epidemics. 2015;10:97-101.

\bibitem{coltart2017ebola}
Coltart CE, Lindsey B, Ghinai I, Johnson AM, Heymann DL.
\newblock The Ebola outbreak, 2013--2016: old lessons for new epidemics.
\newblock Philosophical Transactions of the Royal Society B: Biological Sciences. 2017;372(1721):20160297.

\bibitem{who2023marburg}
{World Health Organisation}. Marburg virus disease -- Equatorial Guinea; 2023.
\newblock Available from: \url{https://www.who.int/emergencies/disease-outbreak-news/item/2023-DON472}.

\bibitem{baker2020successful}
Baker MG, Wilson N, Anglemyer A.
\newblock Successful elimination of Covid-19 transmission in New Zealand.
\newblock New England Journal of Medicine. 2020;383(8):e56.

\bibitem{thompson2019rigorous}
Thompson RN, Morgan OW, Jalava K.
\newblock Rigorous surveillance is necessary for high confidence in end-of-outbreak declarations for Ebola and other infectious diseases.
\newblock Philosophical Transactions of the Royal Society B. 2019;374(1776):20180431.

\bibitem{parag2020exact}
Parag KV, Donnelly CA, Jha R, Thompson RN.
\newblock An exact method for quantifying the reliability of end-of-epidemic declarations in real time.
\newblock PLoS Computational Biology. 2020;16(11):e1008478.

\bibitem{djaafara2021quantitative}
Djaafara BA, Imai N, Hamblion E, Impouma B, Donnelly CA, Cori A.
\newblock A quantitative framework for defining the end of an infectious disease outbreak: application to Ebola virus disease.
\newblock American Journal of Epidemiology. 2021;190(4):642-51.

\bibitem{thompson2024using}
Thompson R, Hart W, Keita M, Fall I, Gueye A, Chamla D, et~al.
\newblock Using real-time modelling to inform the 2017 Ebola outbreak response in DR Congo.
\newblock Nature Communications. 2024;15(1):5667.

\bibitem{nishiura2016objective}
Nishiura H, Miyamatsu Y, Mizumoto K.
\newblock Objective determination of end of MERS outbreak, South Korea, 2015.
\newblock Emerging Infectious Diseases. 2016;22(1):146.

\bibitem{nishiura2016methods}
Nishiura H.
\newblock Methods to determine the end of an infectious disease epidemic: a short review.
\newblock In: Chowell G, Hyman J, editors. Mathematical and statistical modeling for emerging and re-emerging infectious diseases. Springer; 2016. p. 291-301.

\bibitem{hart2024optimizing}
Hart WS, Buckingham JM, Keita M, Ahuka-Mundeke S, Maini PK, Polonsky JA, et~al.
\newblock Optimizing the timing of an end-of-outbreak declaration: Ebola virus disease in the Democratic Republic of the Congo.
\newblock Science Advances. 2024;10(27):eado7576.

\bibitem{bradbury2023exact}
Bradbury NV, Hart WS, Lovell-Read FA, Polonsky JA, Thompson RN.
\newblock Exact calculation of end-of-outbreak probabilities using contact tracing data.
\newblock Journal of the Royal Society Interface. 2023;20(209):20230374.

\bibitem{linton2022epidemics}
Linton NM, Lovell-Read FA, Southall E, Lee H, Akhmetzhanov AR, Thompson RN, et~al.
\newblock When do epidemics end? Scientific insights from mathematical modelling studies.
\newblock Centaurus. 2022;64(1):31-60.

\bibitem{lloyd2005superspreading}
Lloyd-Smith JO, Schreiber SJ, Kopp PE, Getz WM.
\newblock Superspreading and the effect of individual variation on disease emergence.
\newblock Nature. 2005;438(7066):355-9.

\bibitem{thompson2019improved}
Thompson RN, Stockwin JE, van Gaalen RD, Polonsky JA, Kamvar ZN, Demarsh PA, et~al.
\newblock {Improved inference of time-varying reproduction numbers during infectious disease outbreaks}.
\newblock Epidemics. 2019 12;29:100356.

\bibitem{abbott2020estimating}
Abbott S, Hellewell J, Thompson RN, Sherratt K, Gibbs HP, Bosse NI, et~al.
\newblock {Estimating the time-varying reproduction number of SARS-CoV-2 using national and subnational case counts}.
\newblock Wellcome Open Research. 2020 12;5:112.

\bibitem{cori2013new}
Cori A, Ferguson NM, Fraser C, Cauchemez S.
\newblock {A new framework and software to estimate time-varying reproduction numbers during epidemics}.
\newblock American Journal of Epidemiology. 2013 11;178(9):1505-12.

\bibitem{watson2024jointly}
Watson LM, Plank MJ, Armstrong BA, Chapman JR, Hewitt J, Morris H, et~al.
\newblock Jointly estimating epidemiological dynamics of Covid-19 from case and wastewater data in Aotearoa New Zealand.
\newblock Communications Medicine. 2024.

\bibitem{who2018ebola}
{World Health Organization}. Ebola outbreak in DRC ends: WHO calls for international efforts to stop other deadly outbreaks in the country; 2018.
\newblock Available from: \url{https://www.who.int/news-room/detail/24-07-2018-ebola-outbreak-in-drc-ends--who-calls-for-international-efforts-to-stop-other-deadly-outbreaks-in-the-country}.

\bibitem{britton2019estimation}
Britton T, Scalia~Tomba G.
\newblock Estimation in emerging epidemics: biases and remedies.
\newblock Journal of the Royal Society Interface. 2019;16(150):20180670.

\bibitem{charniga2024best}
Charniga K, Park SW, Akhmetzhanov AR, Cori A, Dushoff J, Funk S, et~al.
\newblock Best practices for estimating and reporting epidemiological delay distributions of infectious diseases.
\newblock PLoS Computational Biology. 2024;20(10):e1012520.

\bibitem{gordon1993novel}
Gordon NJ, Salmond DJ, Smith AF.
\newblock Novel approach to nonlinear/non-Gaussian Bayesian state estimation.
\newblock In: IEE Proceedings F (radar and signal processing). vol. 140. IET; 1993. p. 107-13.

\bibitem{sarkka2023bayesian}
S{\"a}rkk{\"a} S.
\newblock Bayesian Filtering and Smoothing.
\newblock Cambridge University Press; 2013.

\bibitem{nz_covid_data}
{New Zealand Ministry of Health}. New Zealand COVID-19 Data; 2024.
\newblock Available from: \url{https://github.com/minhealthnz/nz-covid-data}.

\bibitem{ferretti2020quantifying}
Ferretti L, Wymant C, Kendall M, Zhao L, Nurtay A, Abeler-D{\"o}rner L, et~al.
\newblock Quantifying SARS-CoV-2 transmission suggests epidemic control with digital contact tracing.
\newblock Science. 2020;368(6491):eabb6936.

\bibitem{hendy2021mathematical}
Hendy S, Steyn N, James A, Plank MJ, Hannah K, Binny RN, et~al.
\newblock Mathematical modelling to inform New Zealandâ€™s COVID-19 response.
\newblock Journal of the Royal Society of New Zealand. 2021;51(sup1):S86-S106.

\bibitem{james2021model}
James A, Plank MJ, Hendy S, Binny RN, Lustig A, Steyn N.
\newblock Model-free estimation of COVID-19 transmission dynamics from a complete outbreak.
\newblock PLoS One. 2021;16(3):e0238800.

\bibitem{wegehaupt2023superspreading}
Wegehaupt O, Endo A, Vassall A.
\newblock Superspreading, overdispersion and their implications in the SARS-CoV-2 (COVID-19) pandemic: a systematic review and meta-analysis of the literature.
\newblock BMC Public Health. 2023;23(1):1003.

\bibitem{nkengasong2018response}
Nkengasong JN, Onyebujoh P.
\newblock Response to the Ebola virus disease outbreak in the Democratic Republic of the Congo.
\newblock Lancet. 2018;391(10138):2395-8.

\bibitem{mbala20192018}
Mbala-Kingebeni P, Pratt CB, Wiley MR, Diagne MM, Makiala-Mandanda S, Aziza A, et~al.
\newblock 2018 Ebola virus disease outbreak in {\'E}quateur Province, Democratic Republic of the Congo: a retrospective genomic characterisation.
\newblock Lancet Infectious Diseases. 2019;19(6):641-7.

\bibitem{vankerkhove2015review}
{Van Kerkhove} MD, Bento AI, Mills HL, Ferguson NM, Donnelly CA.
\newblock A review of epidemiological parameters from Ebola outbreaks to inform early public health decision-making.
\newblock Scientific Data. 2015;2(1):1-10.

\bibitem{who2014ebola}
{WHO Ebola Response Team}.
\newblock Ebola virus disease in West Africa â€” the first 9 months of the epidemic and forward projections.
\newblock New England Journal of Medicine. 2014;371(16):1481-95.

\bibitem{eichner2011incubation}
Eichner M, Dowell SF, Firese N.
\newblock Incubation period of Ebola hemorrhagic virus subtype Zaire.
\newblock Osong Public Health and Research Perspectives. 2011;2(1):3-7.

\bibitem{velasquez2015time}
Vel{\'a}squez GE, Aibana O, Ling EJ, Diakite I, Mooring EQ, Murray MB.
\newblock Time from infection to disease and infectiousness for Ebola virus disease, a systematic review.
\newblock Clinical Infectious Diseases. 2015;61(7):1135-40.

\bibitem{althaus2015ebola}
Althaus C.
\newblock Ebola superspreading.
\newblock Lancet Infectious Diseases. 2015;15(5):507-8.

\bibitem{james2007event}
James A, Pitchford JW, Plank MJ.
\newblock An event-based model of superspreading in epidemics.
\newblock Proceedings of the Royal Society B: Biological Sciences. 2007;274(1610):741-7.

\bibitem{dalziel2018unreported}
Dalziel BD, Lau MS, Tiffany A, McClelland A, Zelner J, Bliss JR, et~al.
\newblock Unreported cases in the 2014-2016 Ebola epidemic: Spatiotemporal variation, and implications for estimating transmission.
\newblock PLoS Neglected Tropical Diseases. 2018;12(1):e0006161.

\bibitem{jarvis2022measuring}
Jarvis CI, Gimma A, Finger F, Morris TP, Thompson JA, Le~Polain~de Waroux O, et~al.
\newblock Measuring the unknown: an estimator and simulation study for assessing case reporting during epidemics.
\newblock PLoS Computational Biology. 2022;18(5):e1008800.

\bibitem{gibbons2014measuring}
Gibbons CL, Mangen MJ, Plass D, Havelaar AH, Brooke RJ, Kramarz P, et~al.
\newblock Measuring underreporting and under-ascertainment in infectious disease datasets: a comparison of methods.
\newblock BMC Public Health. 2014;14:1-17.

\bibitem{abbate2016potential}
Abbate JL, Murall CL, Richner H, Althaus CL.
\newblock Potential impact of sexual transmission on Ebola virus epidemiology: Sierra Leone as a case study.
\newblock PLoS Neglected Tropical Diseases. 2016;10(5):e0004676.

\bibitem{lee2019sexual}
Lee H, Nishiura H.
\newblock Sexual transmission and the probability of an end of the Ebola virus disease epidemic.
\newblock Journal of Theoretical Biology. 2019;471:1-12.

\end{thebibliography}


\end{document}